\documentclass{article}

\usepackage{arxiv}

\usepackage[utf8]{inputenc} %
\usepackage[T1]{fontenc}    %
\usepackage{hyperref}       %
\usepackage{url}            %
\usepackage{booktabs}       %
\usepackage{amsfonts}       %
\usepackage{nicefrac}       %
\usepackage{microtype}      %
\usepackage{lipsum}		%
\usepackage{textcomp}
\usepackage{enumitem} %
\usepackage{neuralnetwork}
\usepackage{graphicx}
\usepackage{subcaption}
\usepackage{epsfig}

\newenvironment{leftbox}[1]
{\itemize[
	nosep,
	leftmargin=0pt,
	rightmargin=\dimexpr\textwidth-#1\relax,
	itemindent=\parindent,
	listparindent=\parindent,
	]\item[]\relax}
{\enditemize}

\newcommand\wordcount{
	\immediate\write18{texcount -sub=section \jobname.tex  | grep "Section" | sed -e 's/+.*//' | sed -n \thesection p > 'count.txt'}
	(\input{count.txt}words)}

\usepackage[font={sc, small}]{caption}
\usepackage{lipsum,ragged2e}

\newcommand{\figuredesc}[1]{%
	\begingroup
	\par
	\justifying\small
	\noindent #1
	\par
	\endgroup}

\title{Machine Learning algorithms for Financial Asset Price Forecasting }

\author{
  Philip Ndikum
  \thanks{Preprint. Thanks is given to staff members of the Advanced Research Computing (ARC) group \cite{ARC_2015} at the University of Oxford  for computing support. Additional thanks is given to my family and friends for their support. \textcopyright \; Philip Ndikum $2020$. All rights reserved. No part of this publication or the corresponding proprietary software packages may be reproduced, distributed, or transmitted in any form or by any means without the prior written permission of the publisher.} \\
  University of Oxford\\
  Sa\"{i}d Business School\\
  Park End St, Oxford OX1 1HP \\
  \texttt{philip.ndikum.dipfs19@said.oxford.edu} \\ %
  $31$ March $2020$
}

\usepackage{amsthm}

\theoremstyle{definition}
\newtheorem{definition}{Definition}[section]

\usepackage[
backend=bibtex,
style=numeric, %
citestyle=ieee,
sorting=none,
]{biblatex}
\usepackage{tikz,pgfplots,pgf}
\usetikzlibrary{matrix,shapes,arrows,positioning}
\usepackage{amsmath}

\DeclareMathOperator*{\argmin}{arg\,min}

\usepackage{forest}
\usetikzlibrary{fit,positioning}

\tikzset{
	font=\Large\sffamily\bfseries,
	red arrow/.style={
		midway,red,sloped,fill, minimum height=3cm, single arrow, single arrow head extend=.5cm, single arrow head indent=.25cm,xscale=0.3,yscale=0.15,
		allow upside down
	},
	black arrow/.style 2 args={-stealth, shorten >=#1, shorten <=#2},
	black arrow/.default={1mm}{1mm},
	tree box/.style={draw, rounded corners, inner sep=1em},
	node box/.style={white, draw=black, text=black, rectangle, rounded corners},
}

\usepackage{fancyhdr}
\fancypagestyle{AllPages}{
	\chead{ \centering Machine Learning Algorithms for Financial Asset Price Forecasting. \textcopyright \; Philip Ndikum $2020$. All rights reserved. }
}
\begin{document}
\maketitle

\begin{abstract}
This research paper explores the performance of Machine Learning (ML) algorithms and techniques that can be used for financial asset price forecasting. The prediction and forecasting of asset prices and returns remains one of the most challenging and exciting problems for quantitative finance and practitioners alike. The massive increase in data generated and captured in recent years presents an opportunity to leverage Machine Learning algorithms. This study directly compares and contrasts state-of-the-art implementations of modern Machine Learning algorithms on high performance computing (HPC) infrastructures versus the traditional and highly popular Capital Asset Pricing Model (CAPM) on U.S equities data. The implemented Machine Learning models - trained on time series data for an entire stock universe (in addition to exogenous macroeconomic variables) significantly outperform the CAPM on out-of-sample (OOS) test data. 
\end{abstract}


\section{Introduction - Motivations}
The prediction and forecasting of asset prices in international financial markets remains one of the most challenging and exciting problems for quantitative finance practitioners and academics alike \cite{henrique_bruno_2019, heaton_polson_2016}. Driven by a seismic increase in computing power and data researchers and investment firms point their attention to techniques found in Computer Science namely the promising fields of Data Science, Artificial Intelligence (AI) and Machine Learning (ML). Each day humans generate and capture more than 2.5 quintillion bytes of data. More than $90\%$ of the data generated in recorded human history was created in the last few years and it is estimated that this amount will exceed $40$ Zettabytes or $40$ trillion gigabytes by $2020$ \cite{Dobre_2014, Sivarajah_2016}.
\\ \\
This increase in data presents an enormous opportunity to leverage techniques and algorithms developed in the field of Machine Learning especially it's sub-field Deep Learning. Machine and Deep Learning prediction algorithms have been specifically designed to deal with large volume, high dimensionality and unstructured data making them ideal candidates to solve problems in an enormous number of fields \cite{chen_big_2014, Najafabadi_2015}. Companies around the world have made breakthroughs successfully commercializing Machine Learning R\&D (Research and Development) and notable progress has been made in the fields of medicine, computer vision (in autonomous driving systems) as well as robotics \cite{li2017deep,Obermeyer_2016}. Studies estimate the annual potential value of Machine Learning applied in banking and the financial services sector as much $5.2$ percent of global revenues which approximates to $\$300$bn \cite{Buchanan_2019, Mckinsey_2018}. When contrasted with traditional finance models (which are largely linear) Machine Learning algorithms present the promise of accurate prediction utilising new data sources previously not available. Investment professionals often refer to this non traditional data as ``alternative data" \cite{monk_2018}. Examples of alternative data include the following:
\begin{itemize}
	\item Satellite imagery to monitor economic activity. Example applications: Analysis of spatial car park traffic to aid the forecasting of sales and future cash flows of commercial retailers. Classifying the movement of shipment containers and oil spills for commodity price forecasting \cite{Orfanidis_2018}. Forecasting real estate price directly from satellite imagery \cite{bency_spatial_2017}.
	\item Social-media data streams to forecast equity prices \cite{bollen2011twitter, zhang2011predicting} and potential company acquisitions \cite{Xiang_2012}.
	\item E-commerce and credit card transaction data \cite{2016_butaru} to forecast retail stock prices \cite{Darroch_2017}.
	\item ML algorithms for patent analysis to support the prediction of Merger and Acquisitions (M\&A) \cite{Wei_2009}.
\end{itemize}
Performant Machine Learning algorithm should be able to capture non-linear relationships from a wide variety of data sources. Heaton and Polson \cite{heaton_polson_2016} state the following about Machine Learning algorithms applied to finance:
\begin{quote}
	``Problems in the financial markets may be different from typical deep learning \footnote{For the purposes of simplicity this paper will use the terms Deep Learning and Neural Networks interchangeably.} applications in many respects...In theory [however] a [deep neural network] can find the relationship for a return, no matter how complex and non-linear...This is a far cry from both the simplistic linear factor models of traditional financial economics and from the relatively crude, ad hoc methods of statistical arbitrage and other quantitative asset management techniques" \cite[p.~1]{heaton_polson_2016}
\end{quote}
In recent years a greater number of researchers have demonstrated the impressive empirical performance of Machine Learning algorithms for asset price forecasting when compared with models developed in traditional statistics and finance \cite{gu_2018, chen_2019, RePEc:grz:wpaper:2019-06, RePEc:arx:papers:1909.04497, heaton_2016}. Whilst this paper tests results on U.S equities data, the theory and concepts can be applied to any financial asset class - from real estate, fixed-income \cite{Martn2018MachineLM} and commodities to more exotic derivatives such as weather, energy \cite{Cramer_2017} or cryptocurrency derivatives \cite{2018_Alessandretti, Lahmiri_2019}. The ability of an individual or firm to more accurately estimate the expected price of any asset has enormous value to practitioners in the fields of corporate finance, strategy, private equity in addition to those in the fields of trading and investments. In recent years central banks including Federal Reserve Banks in the U.S \cite{Lemieux_2018} and the Bank of England \cite{Andreas_2019, Chakraborty_2017} have also attempted to leverage Machine Learning techniques for policy analysis and macroeconomic decision making. Before we delve into the world of Machine Learning we shall first lay the ground work of traditional financial theories using the highly popular Capital Asset Pricing Model (CAPM) focusing on the theory from a practitioner's perspective.

\section{The Capital Asset Pricing Model (CAPM)}
\subsection{Introduction}
\label{subsect:CAPM}
The Capital Asset Pricing Model (CAPM) was independently developed in the $1960$'s by William Sharpe \cite{Sharpe_1964}, John Lintner \cite{Lintner_1965}, Jack Treynor \cite{Treynor_1961} and Jan Mossin \cite{Mossin_1966} building on the earlier work of Markovitz and his \textit{mean-variance} and market portfolio models \cite{Markowitz_1959}. In $1990$ Sharpe received a Nobel Prize for his work on the CAPM and the theory remains one of the most widely taught on MBA (Master of Business Administration) and graduate finance and economics courses \cite{Womack_2001}. The CAPM is a one-factor model that assumes the market risk or \textit{beta} is the only relevant metric required to determine a theoretical expected rate of return for any asset or project.

\subsection{Assumptions and Model}
The CAPM holds the following main assumptions:
\begin{enumerate}
	\item One-period investment model: All investors invest over the same one-period time horizon.
	\item Risk averse investors: This assumption was initially developed by Markovitz and asserts that all investors are rational and risk averse actors in the sense that when choosing between financial portfolios investors aim to optimize the following:
		\begin{enumerate}
			\item Minimize the variance of the portfolio returns.
			\item Maximize the expected returns given the variance.
		\end{enumerate}
	\item Zero transaction costs: There are no taxes or transactional costs.
	\item Homogenous information: All investors have homogenous views and information regarding the probability distributions of all security returns.
	\item Risk free rate of interest: All investors can lend and borrow at the at a specified risk free rate of interest $r_f$. 
\end{enumerate}
The CAPM provides an equilibrium relationship \cite{Mossin_1966, Nielsen_1990} between investments $i$'s expected return and its market beta $\beta_i$ and is mathematically defined as follows:
\begin{equation}
\mathbb{E}[r_i] = 
r_f + \beta_i(\mathbb{E}[r_M] - r_f).
\label{eqn:capm}
\end{equation}
If the subscript $M$ denotes the market portfolio then we have:
\begin{align*}
	\mathbb{E}[r_i] &\equiv \text{expected return on asset} \; i \\
	r_f &\equiv \text{risk-free rate of return} \\
	\beta_i &\equiv  \text{cov}(r_i, r_M) / \sigma_M^2 \; \text{is the beta of asset} \; i \\
	\mathbb{E}[r_M] &\equiv \text{expected return on the market portfolio} \\
	\mathbb{E}[r_M] - r_f &\equiv \text{is the market risk premium (MRP)}
\end{align*}
A survey \cite{Graham_2001} of $4,440$ international companies demonstrated that the CAPM is one of the most popular models used by company CFO's (Chief Financial Officers) to calculate the cost of equity. The cost of equity is one of the inputs for the classical Weighted Average Cost of Capital (WACC) of a firm. The pre-tax WACC of a levered firm \cite{Farber_2005} is given by 
\begin{equation}
	\text{WACC} = \frac{D}{V} r_D + \frac{E}{V} r_E
\end{equation}
where $r_D, r_E$ are the cost of equity and cost of debt respectively. $D,E$ and $V$ are the respective values of debt, equity and the net value of a respective firm or project. The WACC has broad array of applications and is often used by practitioners as the \textit{discount rate} in present value calculations that estimate the value of firms in M\&A \cite{Arzac_2004}, individual company projects, as well as options \cite{Arnold_2004} based on forecasted future cash flows. Let us take the example of valuing a company using Discounted Cash Flow (DCF) analysis. The discrete form of the DCF model to value a company can be defined as follows \footnote{The discrete DCF model can be found in most popular graduate finance textbooks such as \cite{1990_Copeland, berk2017corporate}. The $FCF$ can be computed as follows $FCF = EBIT + D\&A - Taxes - CAPEX - \Delta NWC$.}:
\begin{equation}
	\text{Company Value} = \sum_{t=0}^{n} \frac{FCF_t}{(1+r)^t} 
	+ \frac{\text{Terminal Value}}{(1+r)^{n+1}}
\end{equation}
where $FCF_t$ denotes the forecasted future cash flow at year $t$ and $r$ denotes the discount rate which we can let be equivalent to the WACC (let $r=$ WACC). When using DCF models the computed cost of equity from the CAPM and the subsequent WACC calculation will therefore have a significant impact on the estimated company value. 
\subsection{Criticism - Empirical Performance}
The CAPM does not provide the practitioner with any insight on how to correctly apply the model and thus has lead to a broad array of empirical results and corresponding criticisms from researchers such as Fama and French \cite{Fama_2004} in addition to Dempsey \cite{Dempsey_2013} who argue that the model has poor empirical performance and should be abandoned entirely. In relation to technical implementations many researchers (including those in the early literature) relaxed some of the restrictive assumptions of the CAPM to produce impressive empirical results relative to the models simplicity \cite{Blume_1970, Brennan_1971, Jagannathan_1994}.
\\ \\
 In \cite{Brown_2013}, Brown and Terry argue that given the broad CAPM model implementations, it is invalid to make arguments based on the computational evidence and firmly assert that the CAPM or at least one of its variant models will remain in the core finance literature for many years to come. Additionally Partington \cite{Partington_2013} states the following in response to modern researchers criticizing the empirical results of the CAPM: ``Empirical tests of the CAPM probably don’t tell us much one way or another about the validity of the CAPM, but they have revealed quite a lot about correlations between variables in relation to the cross-section of realized returns" \cite{Partington_2013}. Full technical details of model implementation will be explored in sub section \ref{subsec:model_implementations}. Unlike the Data Science and Machine Learning algorithms we will explore in the next section the implementation of the Capital Asset Pricing Model historically can be seen as much of an art form as it is a science. The ML algorithms that are implemented can theoretically be applied to the same use cases as the CAPM - whether that be in expected returns prediction or corporate valuation.
 
\newpage
\section{Machine Learning Algorithms}

Whilst the terms \textit{Machine Learning} and \textit{Artificial Intelligence} are ill-defined in the current literature \cite{Ryll2019EvaluatingTP} we shall use classic definition provided by \cite{marr1977artificial} which defines Artificial Intelligence as the ``isolation of a particular information processing problem, the formulation of a computational theory for it, the construction of an algorithm that implements it, and a practical demonstration that the algorithm is successful". 
\\ \\
In the context of financial asset price forecasting the information processing problem we are trying to solve is the prediction of an asset price $t$ time steps in the future - we are effectively trying to solve a non-linear multivariate time series problem. Our Machine Learning algorithms and techniques should extract patterns learned from historical data in a process known as ``training" and subsequently make accurate predictions on new data. The assessment of the accuracy of our algorithms is known as ``testing" or ``performance evaluation". Whilst there exist a large number of types and classes of Machine Learning algorithms \cite{Ayodele_2010}
a high percentage of the research papers in the current academic literature frame the problem of financial asset price forecasting as a ``supervised learning" problem \cite{yoo_2005, krollner_2010, henrique_bruno_2019, Ryll2019EvaluatingTP}. Given the practical and empirical focus of this paper - strict algorithm definitions and mathematical proofs of algorithms will be omitted - instead a simple theoretical framework of supervised learning is provided in the next subsection.

\subsection{Supervised Learning - theoretical framework}
\label{subsect:supervised_learning}
\begin{definition}[Supervised Learning]
	Given a set of $N$ example input-output pairs of data 
	\begin{equation*}
	\mathcal{D} = (\mathbf{x}_1,y_1), (\mathbf{x}_2,y_2), \cdots, (\mathbf{x}_n,y_n)
	\end{equation*}
	we first assume that $y$ was generated by some unknown function $f(\mathbf{x})=y$ which can be modelled by our supervised learning algorithm (the ``learner"). $\mathbf{x}_n = (x_1, x_2, \cdots, x_j)$ here denotes a vector containing the explanatory input variables and $y$ is the corresponding \textit{response}. If we are doing a batch prediction on multiple input vectors (denoted by the matrix $X$) we can denote our mapping using $f(X) \to \mathbf{y}$ where $\mathbf{y} = (y_1, y_2, \cdots, y_n)$. In general we decompose our data $\mathcal{D}$ into a training set and a test set. In the training phase our algorithm will learn to approximate our function to produce a prediction $\hat{y}$ - we will denote the approximated function as $\hat{f}(\mathbf{x})$. In Machine Learning we evaluate the performance of algorithm using a accuracy measure which is usually a function of the error terms in the test set - we will denote this performance metric as $p(\hat{y} - y)$. The standard formulation of supervised learning tasks are known as ``classification" and ``regression":
	\begin{itemize}
		\item Classification: The learner is required classify the probability of an event or multiple events occurring. Thus we have the mapping $\hat{f}:\mathbf{x} \to \hat{y}$ such that $\hat{y} \in [0,1]$. Examples: Classifying the probability of an economic recession happening for a target nation or classifying the probability of a target company being merged or acquired (M\&A prediction). 
		\item Regression: The learner is required to predict a real number as the output. Thus we have the mapping $\hat{f}:\mathbf{x} \to \hat{y} \in \mathbb{R}$. Examples: Predicting the annual returns of a financial asset $t$ time periods in the future (which will be implemented in this paper). Another example is the forecasting of interest rates or yield curves.
	\end{itemize}
\end{definition}
Independent of the algorithm or class of algorithms that are selected and implemented, the function $\hat{f}$ is usually estimated using techniques and algorithms from the fields of Applied Mathematics known as \textit{Numerical Analysis and Optimization}. Bennett and Parrado-Hernandez \cite[p.~1266]{bennett_2006} make the important observation that ``optimization lies at the heart of machine learning. Most machine learning problems reduce to optimization problems". In supervised learning we will determine our function $\hat{f}$ by iteratively minimizing a loss function $\mathcal{L}(f, y)$ which minimizes the error on the parsed training data \cite{ng_2011}. Mathematically in a general form we have
\begin{equation}
\hat{f}(X) = \argmin_{f(X)} \mathcal{L}(f(X), y)
\label{eqn:loss_function}
\end{equation}
which is an optimization problem. Independent of the algorithm or loss function choice the ML literature is heavily concerned with addition of a term to equation \ref{eqn:loss_function} known as a regularization term \cite{zaremba_2014, 2004_ng}. This is a mathematical term used to reduce the problem of \textit{overfitting} \cite{2004_hawkins}. Backtest overfitting in finance refers to models which perform well on historical data but ultimately do not perform well ``in the wild" (on new live data streams). An additional benefit of ML techniques for financial asset pricing is that modern researchers are heavily focussed on designing algorithms and techniques that systematically reduce the problem of overfitting to increase the probability of accurate forecasts on new data streams \cite{Salehipour_2016, bailey_2017}.

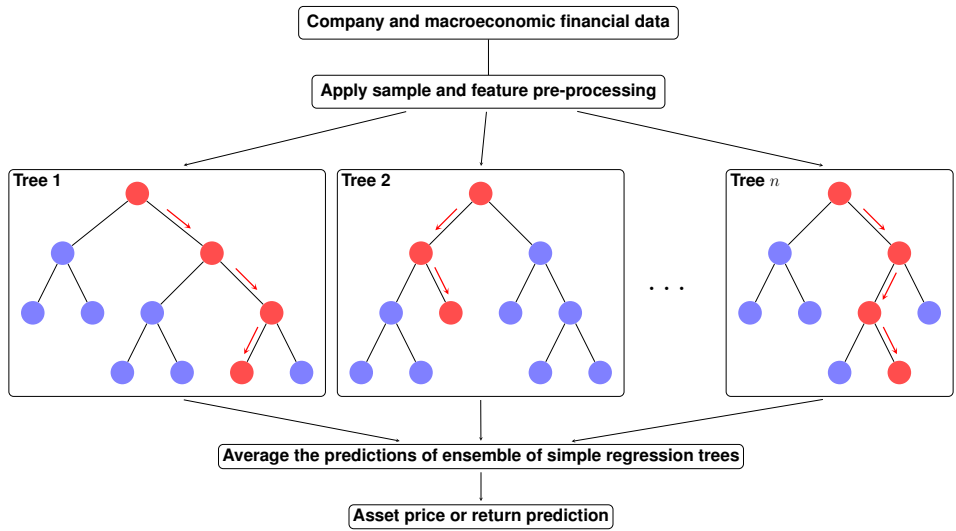
\begin{figure}[htp!]
	\centering
	\caption{Example of Gradient Boosted Tree architecture}
	\scalebox{0.45}{
		\begin{forest}
			for tree={l sep=3em, s sep=3em, anchor=center, inner sep=0.7em, fill=blue!50, circle, where level=2{no edge}{}}
			[
			Company and macroeconomic financial data, node box
			[Apply sample and feature pre-processing, node box, alias=bagging, above=4em
			[,red!70,alias=a1[[,alias=a2][]][,red!70,edge label={node[above=1ex,red arrow]{}}[[][]][,red!70,edge label={node[above=1ex,red arrow]{}}[,red!70,edge label={node[below=1ex,red arrow]{}}][,alias=a3]]]]
			[,red!70,alias=b1[,red!70,edge label={node[below=1ex,red arrow]{}}[[,alias=b2][]][,red!70,edge label={node[above=1ex,red arrow]{}}]][[][[][,alias=b3]]]]
			[~~$\dots$~,scale=2,no edge,fill=none,yshift=-4em]
			[,red!70,alias=c1[[,alias=c2][]][,red!70,edge label={node[above=1ex,red arrow]{}}[,red!70,edge label={node[above=1ex,red arrow]{}}[,alias=c3][,red!70,edge label={node[above=1ex,red arrow]{}}]][,alias=c4]]]]
			]
			\node[tree box, fit=(a1)(a2)(a3)](t1){};
			\node[tree box, fit=(b1)(b2)(b3)](t2){};
			\node[tree box, fit=(c1)(c2)(c3)(c4)](tn){};
			\node[below right=0.5em, inner sep=0pt] at (t1.north west) {Tree 1};
			\node[below right=0.5em, inner sep=0pt] at (t2.north west) {Tree 2};
			\node[below right=0.5em, inner sep=0pt] at (tn.north west) {Tree $n$};
			\path (t1.south west)--(tn.south east) node[midway,below=4em, node box] (mean) {Average the predictions of ensemble of simple regression trees};
			\node[below=3em of mean, node box] (pred) {Asset price or return prediction};
			\draw[black arrow={5mm}{4mm}] (bagging) -- (t1.north);
			\draw[black arrow] (bagging) -- (t2.north);
			\draw[black arrow={5mm}{4mm}] (bagging) -- (tn.north);
			\draw[black arrow={5mm}{5mm}] (t1.south) -- (mean);
			\draw[black arrow] (t2.south) -- (mean);
			\draw[black arrow={5mm}{5mm}] (tn.south) -- (mean);
			\draw[black arrow] (mean) -- (pred);
		\end{forest}
	}
	\figuredesc{Gradient boosting tree algorithms are greedy algorithms that first pre-process the data in a process known as sample and feature bagging \cite{Friedman_2001}. In contrast to the CAPM which is a single model gradient boosted tree algorithms train an \emph{ensemble} of weak base learners (simple regression trees). These base learners are aggregated through a function such as an average to produce the final asset price prediction. The final function is estimated through the minimization of the pre-defined loss function described in \ref{subsect:supervised_learning}.}
	\label{fig:random_forests}
\end{figure} 

\begin{figure}
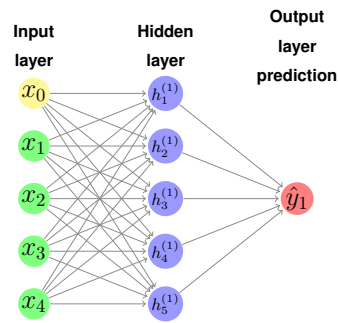

	\centering
	\caption{Example Feed-Forward Neural Network Architectures}
	\begin{subfigure}[b]{0.55\textwidth}
		\centering
		\begin{leftbox}{11cm}
			\scalebox{0.7}{
				\begin{neuralnetwork}[height=5]
					\newcommand{\x}[2]{$x_#2$}
					\newcommand{\y}[2]{$\hat{y}_#2$}
					\newcommand{\hfirst}[2]{\small $h^{(1)}_#2$}
					\newcommand{\hsecond}[2]{\small $h^{(2)}_#2$}
					\inputlayer[count=4, bias=true, title=\small Input\\layer, text=\x]
					\hiddenlayer[count=5, bias=false, title=\small Hidden\\layer , text=\hfirst] \linklayers
				
					\outputlayer[count=1, title= \small Output\\layer prediction, text=\y] \linklayers
				\end{neuralnetwork}
			}
		\end{leftbox}
		\caption{Shallow Feed-Forward Neural Network Architecture}
		\label{fig:sub1}
	\end{subfigure}%
	
	\begin{subfigure}[b]{0.55\textwidth}
		\begin{leftbox}{12cm}
			\scalebox{0.7}{
				\begin{neuralnetwork}[height=5]
					\newcommand{\x}[2]{$x_#2$}
					\newcommand{\y}[2]{$\hat{y}_#2$}
					\newcommand{\hfirst}[2]{\small $h^{(1)}_#2$}
					\newcommand{\hsecond}[2]{\small $h^{(2)}_#2$}
					\inputlayer[count=3, bias=true, title=\small Input\\layer, text=\x]
					\hiddenlayer[count=4, bias=false, title=\small Hidden\\layer 1, text=\hfirst] \linklayers
					\hiddenlayer[count=5, bias=false, title=\small Hidden\\layer 2, text=\hsecond] \linklayers
					\hiddenlayer[count=5, bias=false, title=\small Hidden\\layer 3, text=\hsecond] \linklayers
					\hiddenlayer[count=5, bias=false, title=\small Hidden\\layer 4, text=\hsecond] \linklayers
					\hiddenlayer[count=2, bias=false, title=\small Hidden\\layer 5, text=\hsecond] \linklayers
					\outputlayer[count=1, title= \small Output\\layer prediction, text=\y] \linklayers
				\end{neuralnetwork}
			}
		\end{leftbox}
		\caption{Deep Feed-Forward Neural Network Architecture}
	\end{subfigure}
	\label{fig:neural_network}
	\figuredesc{The figures above demonstrate example neural network architectures for algorithms known as feed-forward neural networks (FNN). These algorithms were originally inspired by the structure of the human brain \cite{Robert_1988}: Each neural network is composed of nodes and hidden layers which undertake individual mathematical operations to form a computational graph. In accordance with the universal approximation theorem \cite{Cybenko1989ApproximationBS}, neural networks are stated to have the potential to be universal approximators meaning they can theoretically approximate any mathematical function provided the appropriate architecture. The process of determining this function is also determined through the minimization of a pre-defined loss function described in sub-section \ref{subsect:supervised_learning}. Although this paper implements FNN's for prediction, a hybrid of FNN, convolutional neural networks (CNN) and recurrent neural networks (RNN) architectures have also proved to be performant for asset price and return forecasting in recent papers \cite{Stoean_2019, Persio_2016, Sirignano_2019}.}
\end{figure}

\subsection{Incorporating regulatory and financial constraints into ML algorithms}
 \label{subsect:regulation}
 In financial asset pricing (and quantitative finance more broadly) professionals are not only concerned with accurate forecasts but are also constrained by investor risk appetite, traditional econometric and financial theory (such as CAPM), and more increasingly - international regulatory concerns in North America and Europe \cite{2017_Cath}. Many in the investment world have been sceptical about Machine Learning due to misconceptions that the algorithms are closed sourced and black boxes \cite{1997_benitez, wang_2007}. One can argue that this scepticism is warranted - \cite{chiu_2016} notes that the global financial crisis of $2007-2009$ caused regulators to move away from an excessively laissez-faire approach to financial regulation to an aggressive, forward looking and proactive approach. Additionally from a psychological perspective, Dietvorst et al \cite{Dietvorst_2014} have shown even though algorithmic forecasts outperform human forecasters by at least $10\%$ across multiple domains a body of Psychology research demonstrates that humans have a low tolerance to the errors of machines and algorithms - a phenomenon coined as ``algorithm aversion". As new technologies and algorithms develop regulators have therefore been quick to introduce new policies and laws to ensure adequate protections to society and the general public \cite{Kirilenko_2013}. Regulators from the European Union (E.U) have acted swiftly in their implementation of a large number of laws such as the E.U General Data Protection Regulation (GDPR) and the Markets in Financial Instruments Directive (MiFID) II which both came into effect in $2018$. Sheridan and Iain \cite[p.~420]{Shredian_2017} state the following: 
\begin{quote}
	``Under Article 17(2) of MiFID II, a competent authority may require the investment firm to provide, on a regular or ad hoc basis, a description of the nature of its algorithmic strategies, details of its trading parameters or limits to which the system is subject" \cite[p.~420]{Shredian_2017}.
\end{quote}
Additionally Recital 71 of the GDPR affords consumers the rights to ask private institutions to explain ML algorithms \cite{goodman_2016, kush_2016}. Much of the recent innovation in the field of ML for asset pricing directly relates to creating more transparent algorithms that can be explained to both regulators and investors \cite{kou_2019}. Table \ref{table:investor_constraints} provides a high level overview of regulatory constraints and recent ML literature that attempts to solve regulatory problems. We implemented modern statistical techniques to provide explainability to our Machine Learning models. As we move into an increasingly regulated world, state-of-the-art financial asset price forecasting performance and adoption will require the collaboration of experts in the legal, financial and scientific communities.

\begin{table}[htp!]
	\centering
	\small
	\caption{Regulatory constraints relevant to ML for financial asset price forecasting}
	\label{table:investor_constraints}
	\begin{tabular}{ |p{5cm}|p{10cm}| } 
		\hline
		\textbf{Regulatory constraints} & \textbf{Relevant literature solutions}  \\ \hline
		Transparency and \newline Explainability  \cite{goodman_2016, kush_2016,kou_2019, Johnson_2019, Citron_2014}.
		& 
		Finance practitioners and academics have focussed on modifying ML algorithm loss functions to incorporate traditional finance theory to allow for greater transparency. \cite{chen_2019, pelger_2019} for example demonstrate an improvement of algorithm performance when they incorporate no-arbitrage pricing theory constraints from traditional finance \cite{ross_1976}. Additionally \cite{Feng_2017, kelly_2017} systematically utilise modern ML and statistical algorithms to reduce the number of features or ``factors" used in asset pricing - this ultimately serves to improve model transparency in the face of high dimensionality and large unstructured data. 
		\newline \newline
		In the ML literature there has been a drive to develop domain agnostic software packages and tools to explain trained ML models \cite{2018_Adadi}. Most notably techniques such as LIME \cite{ribeiro_2016} and SHAP \cite{Lundberg_2017} have been implemented in multiple popular programming languages to allow for detailed explanations of any supervised learning model.
		\\ \hline
		Risk Management \cite{Schmaltz_2013, Johnson_2018, Ang_2011, Shredian_2017}.
		& There have been a broad array of solutions relating to risk management for ML in the finance literature: \cite{Chandrinos_2018, Aziz_2018} explore how we can develop and incorporate supplementary ML models to statistically account for financial risk in our investment and trading systems. Recent papers such as \cite{kou_2019, Coulombe2019HowIM, Berge_2013} explore how ML can be used to directly forecast macroeconomic variables to identify systematic risk and economic recessions. \cite{Alberg_2017} also demonstrates how transparent models can be built by forecasting company fundamentals (cash flow, and balance sheet line items) rather than asset prices directly.
		\newline \newline
		In the ML literature there has also been a movement away from the development of point forecasts models to Bayesian techniques which allow for probabilistic forecasts \cite{gal_2015, Zhu_2017,Duan2019NGBoostNG, Ghahramani_2015}. These Bayesian algorithms inherently allow the end user to account for risk through posteriori probability distributions: \cite{Spiegeleer_2018,Ruf_2019} review how these Bayesian techniques have been utilized for option pricing and derivatives hedging strategies. 
		\\ \hline
	\end{tabular}
\end{table}

\newpage
\section{Empirical performance on U.S equities}
\label{section:perform_eval}
\subsection{Machine Learning algorithms}
Empirical studies on the best supervised learning algorithms tend to suggest that decision tree and neural network algorithms perform the best across multiple domains and data-sets \cite{caruana_2008, caruana2006empirical}. We will evaluate our algorithms on publicly traded U.S equities data. Researchers evaluating Machine Learning algorithms empirically in the domain of equity price and return forecasting have stressed the importance of both Artificial Neural Network (ANN) architectures and ensembles of decision trees (such as random forests, and gradient boosted trees) \cite{dongdog_2019, krollner_2010, chen_2019, kelly_2017}. For this reason algorithm testing was focussed on Python implementations of neural network, and gradient boosted trees. The gradient boosting tree and neural network architectures are illustrated in Figures \ref{fig:random_forests} and \ref{fig:neural_network} respectively. Modern packages such as NGBoost developed by Duan et al ($2019$) \cite{Duan2019NGBoostNG} which allows for probabilistic forecasting were also implemented and tested.
\\ \\
Each of our Machine Learning algorithms have a unique set of initial parameters known as \emph{hyperparameters}. These hyperparameters are initial model conditions which determine the performance our models. The ideal hyperparameters for optimal model performance cannot be known a priori and thus traditionally would require extensive manual tuning and configuration. In a relatively new sub-field of Machine Learning known as Automated Machine Learning (AutoML) \cite{Yao_2018} researchers have worked hard to develop and create software packages to automate the process of manual hyperparameter optimization using advanced Bayesian and biologically-inspired algorithms \cite{Claesen_2015}. The researcher and practitioner must only defined the search space for each of the model parameters and allow the algorithm to run for $n$ number of trials or simulations, the algorithm will then attempt to search for the hyperparameters that produce the best model performance. Since we are attempting to determine whether Machine Learning algorithms outperform the CAPM model a combination of grid-search and Bayesian hyperparameter optimization was used to determine the best neural network and gradient boosting models. The models were tested using the University of Oxford's Advance Research Computing (ARC) multi-GPU (Graphical Processing Unit) clusters. Table \ref{table:ml_model_parameters} shows a high level overview of the implemented models and their respective optimized hyperparameters. A single model was built for each algorithm to predict the annual returns of our entire stock universe described in the next sub-section. Neural networks were implemented using Keras \cite{chollet2015keras}.
\begin{table}[htp!]
	\renewcommand{\arraystretch}{1.3}
	\caption{Overview of the implemented Machine Learning models and optimized hyperparameters}
	\label{tab:example}
	\centering
	\small
	\begin{tabular}{|p{4cm}|p{4.2cm}| p{7cm} |}
		\hline
		ML Algorithm  &  Optimization Algorithms & Optimized Hyperparameters \\
		\hline
		\hline
		
		NGBoost \cite{Duan2019NGBoostNG}.    &   Grid Search. &  Number of tree estimators (weak base learners).
		\\ 
		\hline
		
		XGBoost.   &    HyperOpt implementation of the Tree of Parzen Algorithm for $n=50$ trials \cite{Bergstra_2013}.
		& Number of tree estimators, maximum depth for each tree, learning rate, regularization parameters, data sampling parameters. 
		\\
		\hline
		
		Catboost \cite{Prokhorenkova_2018}.   &    HyperOpt implementation of the Tree of Parzen Algorithm for $n=50$ trials \cite{Bergstra_2013}. 
		& Number of tree estimators, maximum depth for each tree, learning rate, regularization parameters, data sampling parameters. 
		\\
		\hline
		
		LightGBM \cite{XGboost_2016}.   
		&   HyperOpt implementation of the Tree of Parzen Algorithm for $n=50$ trials \cite{Bergstra_2013}.  
		& Number of tree estimators, maximum depth for each tree, learning rate, regularization parameters, data sampling parameters. 
		\\
		\hline
		
		Shallow Feed-Forward \newline Neural Network (Shallow FNN).  \cite{chollet2015keras}  &   HyperOpt implementation of the Tree of Parzen Algorithm for $n=100$ trials \cite{Bergstra_2013}.
		& 
		Number of hidden layers where $n \in [1,2]$, number of nodes per hidden layer where $n \in [256, 1024]$, 
		Batch normalization configurations for each hidden layer, regularization parameters, activation function configurations for each hidden layer. \newline
		\\
		\hline
		
		Deep Feed-Forward Neural Network (Deep FNN)  \cite{chollet2015keras}.  & HyperOpt implementation of Tree Parzen Algorithm for $n=100$ trials \cite{Bergstra_2013}. 
		& Number of hidden layers where $n \in [3,5]$, number of nodes per hidden layer where $n \in [256, 1024]$, 
		Batch normalization configurations for each hidden layer, regularization parameters, activation function configurations for each hidden layer. \newline
		\\
		\hline
	\end{tabular}
	\label{table:ml_model_parameters}
\end{table}

\subsection{Data and experimental details}
\label{subsec:model_implementations}
\begin{figure}[htp!]
	\centering
	\caption{Example macroeconomic time-series data used in Machine Learning models}
	\scalebox{.63}	{
	\includegraphics{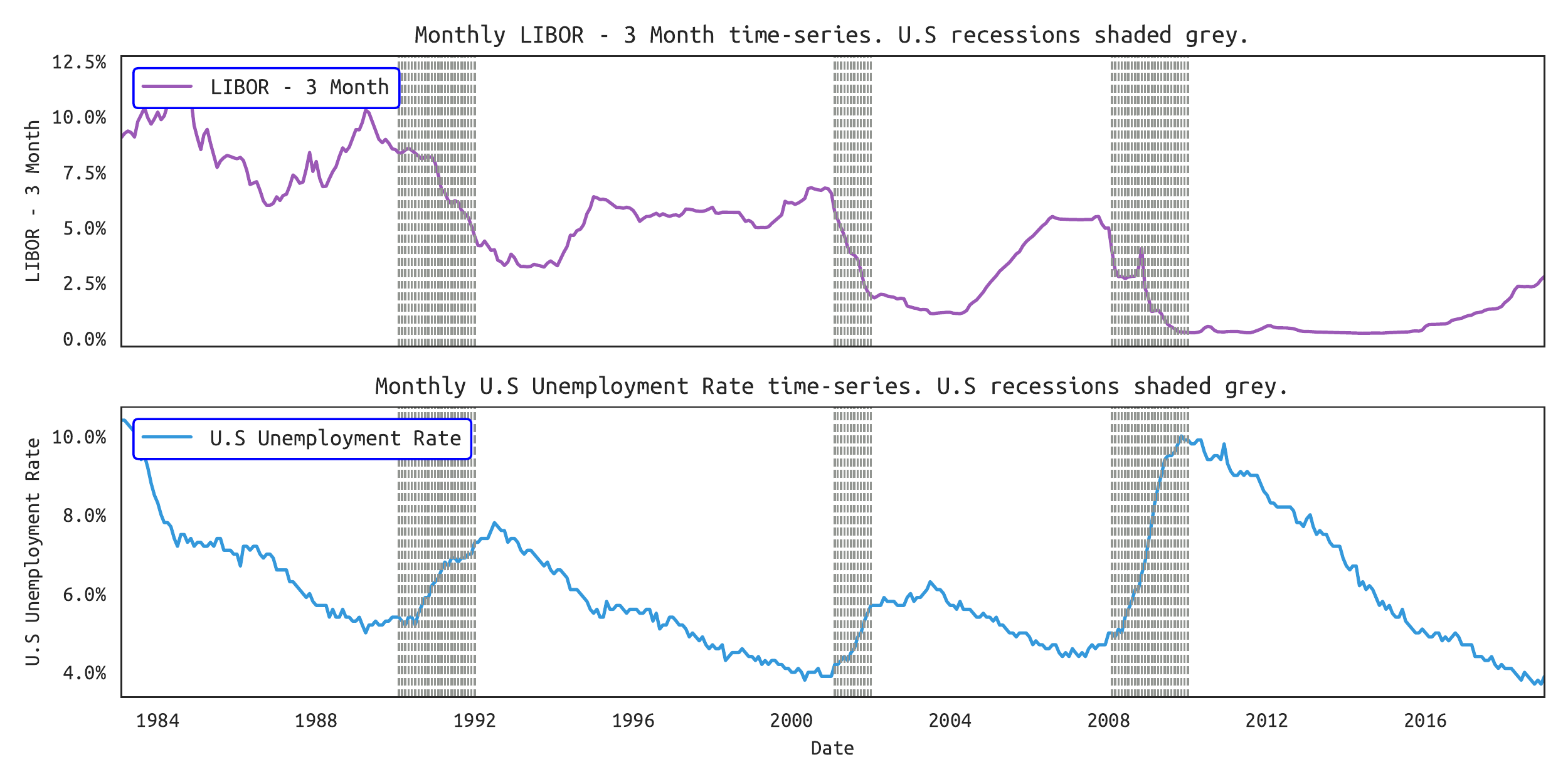}
	}
	\label{fig:macro_ts}
\end{figure}

We conduct a large scale empirical analysis for all publicly traded U.S stocks available on the Wharton Research Data Services (WRDS) cloud \cite{WRDS_1993} that have existed and survived from $1983$ to the start of $2019$ covering a time horizon of $30$ years - this equated to $782$ stocks. The study attempts to predict the \emph{annual returns} of this stock universe using the Machine Learning algorithms shown in Table \ref{table:ml_model_parameters} versus the CAPM. All data was pulled from the WRDS. From the WRDS cloud, data was pulled from the Center for Research in Security Prices (CRSP) and the Standard \& Poor’s (S\&P) Global Market Intelligence Compustat databases. Data was extracted for monthly and annually asset prices, accounting financial statements (balance sheet, income statement, cash flow statement) in addition to macroeconomic factors. Figure \ref{fig:macro_ts} shows a sample of the U.S macroeconomic time series features which included monthly data on consumer price indices, bond rates, gross domestic product (GDP) and other features. 

 A proprietary python software package was built on top of the official WRDS python API's to automate the extraction and transformation of asset price data from WRDS, in addition to the training and testing of both Machine Learning and CAPM models to allow for replicable results. In our computation of the CAPM model shown in equation \ref{eqn:capm} we follow the recommendations of \cite{2014_Plyakha, Pae_2015} and use a value-weighted (VW) U.S S\&P 500 index to represent the market portolio versus an equally weighted (EW) index. Based on empirical studies of CAPM applied to U.S markets $10$ year U.S Treasury Bill returns were used as a proxy for the risk free rate \cite{Mukherji_2011}. Additionally \cite{Chervany_1980, Daves_2000} recommend using a time horizon of roughly three to eight years to estimate the asset beta - a time series window of three years was for all our historical return calculations \footnote{Beta was computed using from 3-year monthly stock and market returns using $\beta_i \equiv  \text{cov}(r_i, r_M) / \sigma_M^2$ }. Based on the literature \cite{Cooper_1996, Jacquier_2003} a simple arithmetic was used versus the geometric mean to compute the annualized average rate of return from monthly returns \footnote{Modern researchers such as Liu et al \cite{Liu_2019} in addition to Hasler and Martineau \cite{Hasler_2019} have argued based on empirical evidence that it is better to compute the annualized monthly return from the arithmetic mean of daily returns. To avoid data sparsity issues over the long time horizon a more conventional approach was followed here.} data over the three year time horizon. 
\\ \\
For our Machine Learning models after data extraction and pre-processing is completed we must generate the training and test sets described in \ref{subsect:supervised_learning}. In time series problems one must maintain the temporal order in the training and test set to prevent what is known as \textit{feature leakage} in the Machine Learning literature or an equivalent term is \textit{look-ahead bias} in finance and econometrics. Assuming we maintain the temporal order of our time series data the canonical approach taken the literature is to use an OOS (out-of-sample) evaluation whereby a sub-sample at the end of the time-series is held out for validation. We will use the phrase \textit{sequential evaluation} to denote this type of validation.
\\ \\
It is important to note that recent researchers \cite{Bergmeir_2012, Roberts_2016} have focused their attention on new forms of \emph{blocked cross-validation} specifically for time-series problems to ensure robust models which do not over-fit datasets. Given the seasonality of annual international stock returns demonstrated by researchers such as Heston and Sadka \cite{Heston_2010} these new methods of evaluating time series forecasts may be of significant interest to practitioners and industrial researchers. Recent empirical studies conducted by Cerqueira et al ($2019$) \cite{Cerqueira_2019} and Mozetic et al \cite{Mozetic_2018} ($2018$) have not demonstrated a significant improvement of these new cross-validation techniques for non-stationary time series data and thus a conventional \textit{sequential evaluation} approach was followed here. 30$\%$ of the data at the end of the time-series was held out for testing reflecting approximately $6$ years of unseen data from $2012-2018$. The final Machine Learning model training and test data-sets contained approximately $200$ features relating to company financial performance in addition to exogenous U.S macroeconomic variables for the three years prior to the prediction year.

\subsection{Results}
\begin{figure}[htp!]
	\centering
	\caption[Feature importance plots]{Example Top 10 feature importance plots for model interpretability created using SHAP (SHapley Additive exPlanations) by Lundberg and Lee (2017) \cite{Lundberg_2017}.}
	\begin{subfigure}[b]{0.55\textwidth}
		\begin{leftbox}{14cm}
		\scalebox{.58}	{
			\includegraphics{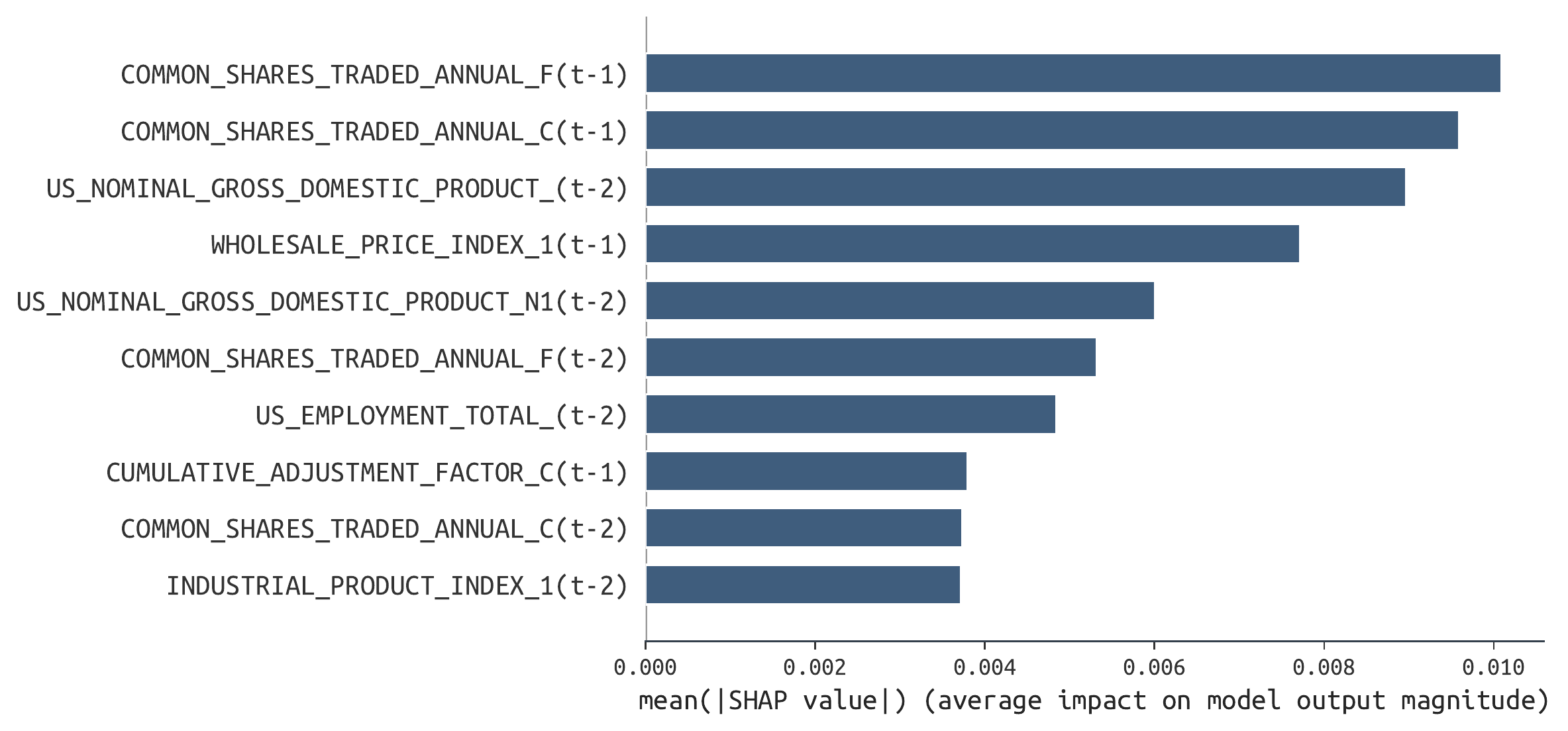}
		}
		\caption{Catboost SHAP Feature Importance Plot}
		\label{fig:catboost} 
	\end{leftbox}
	\end{subfigure}
	
	\begin{subfigure}[b]{0.55\textwidth}
		\begin{leftbox}{14cm}
		\scalebox{.58}	{
			\includegraphics{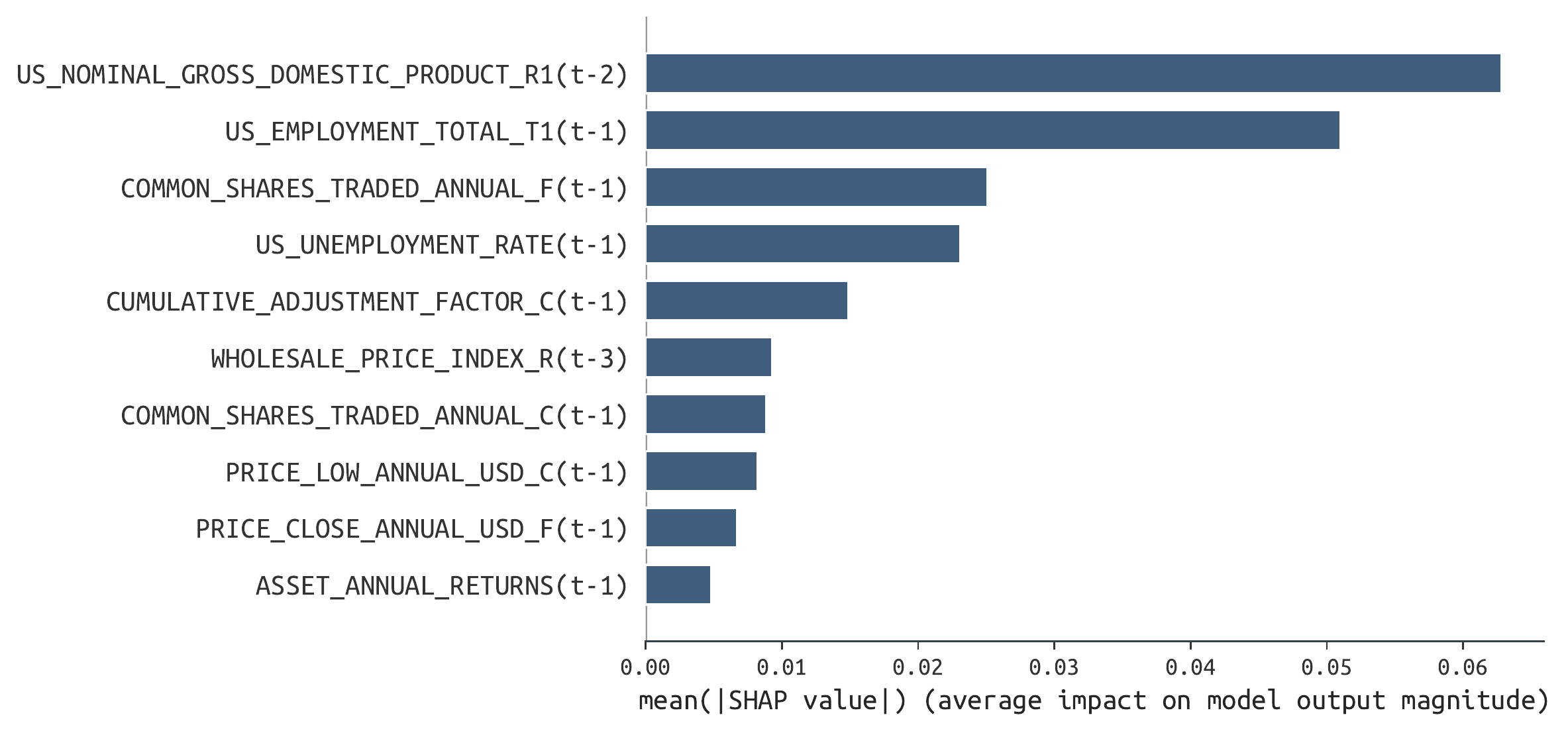}
		}
		\caption{XGBoost SHAP Feature Importance Plot}
		\label{fig:xgboost}
	\end{leftbox}
	\end{subfigure}
	\label{fig:feature_importance_plots}	
\end{figure}

For the performance metric we used the Mean Squared Error (MSE). Given the predicted annual return $\hat{y}$ and the actual annual return $y$ the MSE is defined as follows\footnote{The ideal model is that which minimizes the MSE on OOS (out-of-sample) data.}: 
\begin{equation}
	\text{MSE}(\hat{y} - y) = \frac{1}{n} \sum_{i=1}^{n} (\hat{y}_i - y_i)^2.
\end{equation}
The model results are summarized in Table \ref{tab:example}. The results demonstrate that the Machine Learning techniques result a significant performance improvement over the classical Capital Asset Pricing Model. This demonstrates the power and flexibility of Machine Learning techniques for econometric and financial markets prediction. In line with the literature the gradient boosting tree models performed similarly to the neural network models. Given that the Deep FNN performed better than the Shallow FNN it may be exploring convolutional neural network and recurrent neural network architectures in future work (these architectures have been shown to be highly performant on complex and high dimensionality financial forecasting problems). As computing power, availability of datasets, and scientific innovation continues to increase new model architectures and algorithms will be developed which will ultimately allow the performance of these models to improve with time.
\begin{table}[htp!]
	\renewcommand{\arraystretch}{1.3}
	\caption{Model Results}
	\label{tab:example}
	\centering
	\begin{tabular}{|p{5cm}|p{5cm}| }
		\hline
		Optimized Model  &  Mean Squared Error (MSE)\\
		\hline
		\hline
		
		CAPM   &   $1.6001$\\
		\hline
		
		NGBoost    &   $0.3572$\\
		\hline
		
		XGBoost    &   $0.3280$\\
		\hline
		
		Catboost    &   $0.3125$\\
		\hline
		
		LightGBM    &   $0.3131$\\
		\hline
				
		Shallow FNN    &   $0.3628$\\
		\hline
		
		Deep FNN    &   $0.3531$\\
		\hline
	\end{tabular}
	\figuredesc{}
\end{table}
These results demonstrate the benefit of Machine Learning for financial institutions and practitioners around the world. Whilst we focussed on U.S. equities the same techniques and algorithms could be applied to any asset class. In sub-section \ref{subsect:regulation} we explained the development of packages such as \emph{SHAP} and \emph{LIME} that enable researchers to explain the results of Machine Learning models to investors and regulators. Figures \ref{fig:catboost} and \ref{fig:xgboost} provide the top $10$ features\footnote{Full feature descriptions can be found on the WRDS cloud.} produced by a Python implementation of SHAP (SHapley Additive exPlanations) for the Catboost and XGBoost models. The feature importance plots demonstrate the significance of macroeconomic variables in the prediction of annual returns for U.S equities. As one would expect the U.S GDP, in addition to wholesale and industrial price indices had a significant impact on the predicted returns. One interesting thing to observe is that outside of stock price and stock volume these macroeconomic variables seemed to be far more important than individual stock accounting fundamentals (cash flow statement and balance sheet line items) for the prediction of annual returns in the years $2012-2018$. These trained universal Machine Learning models should generalize to predict annual returns for all U.S equities on future data.  

\subsection{Conclusion}
In conclusion this paper first provided explanations and theoretical frameworks for the CAPM and supervised learning used in Machine Learning. Secondly, this paper explored the regulatory and financial constraints placed on those applying Data Science and Machine Learning techniques in the field of quantitative finance. It was shown that practitioners have a numerous amount of regulatory challenges and hurdles to ensure that scientific innovations can be lawfully adopted and implemented. Finally, we provided a large scale empirical analysis of Machine Learning algorithms versus the CAPM when predicting annual returns of $782$ U.S equities using state-of-the-art Machine Learning techniques and high performance computing (HPC) infrastructures. The results demonstrated the superior performance and power of Machine Learning techniques to forecast annual returns. In contrast to traditional finance theories such as the CAPM, the Machine Learning algorithms had the flexibility to incorporate approximately $200$ time series features to predict the returns for each target U.S equity. As we move further into the age of big data, Data Science and Machine Learning algorithms will increasingly dominate the world of economics and finance. 

\newpage
\setlength{\baselineskip}{0pt} %

{\renewcommand\MakeUppercase[1]{#1}%

\printbibliography[heading=bibintoc,title={References}]

\end{document}